\documentclass[aps,pra,reprint,superscriptaddress,longbibliography]{revtex4-2}

\usepackage{soul}

\usepackage{graphicx}

\usepackage{amsmath,amssymb,amsfonts}

\usepackage{enumitem}

\usepackage{xcolor}

\usepackage{soul,xcolor}

\usepackage{tikz,tkz-euclide}

\usepackage[colorlinks=true,citecolor=blue,linkcolor=blue,urlcolor=blue]{hyperref}

\newcommand{\mean}[1]{\left\langle #1 \right\rangle}

\newcommand{\abs}[1]{\left|#1\right|}

\let\ni\noindent
\usepackage{physics}
\usepackage{verbatim}

\usepackage{tikz,ifthen} 
\usetikzlibrary{shapes,arrows,positioning,automata,backgrounds,calc,er,patterns,matrix}
\newcommand{\Mod}[1]{\ \mathrm{mod}\ #1}
\newcommand{\proj}[1]{\ket{#1}\!\!\bra{#1}}

\begin{document}

\begin{abstract}
    Gauge theories arise in  physical systems displaying space-time local symmetries.  They provide a powerful description of important realms of physics ranging from fundamental interactions, to statistical mechanics, condensed matter and more recently quantum computation. As such, a remarkably deep understanding has been achieved in the field.  With the advent  of quantum technology, lower energy analogs,  capable to capture important features of the original quantum field theories  through quantum simulation, have been intensively studied. Here, we propose a specific scheme implementing an {\it analogic} quantum simulation of lattice gauge theories constrained to mesoscopic spatial scales. To this end, we study the dynamics of mesons residing in a ring-shaped lattice of mesoscopic size pierced by an effective magnetic field. In particular, we find  a new  type of Aharonov-Bohm effect that goes beyond the  particle-like effect and reflecting the the features of the confining gauge potential.
    The coherence properties of the meson are quantified by the persistent current and by specific features of the correlation functions. When the magnetic field is quenched, Aharonov-Bohm oscillations and correlations start a specific matter-wave current dynamics.
\end{abstract}

\title{Aharonov-Bohm effect for confined matter in lattice gauge theories}

\author{Enrico C. Domanti}

\affiliation{Quantum Research Centre, Technology Innovation Institute, Abu Dhabi, UAE}

\affiliation{Dipartimento di Fisica e Astronomia, Via S. Sofia 64, 95127 Catania, Italy}

\affiliation{INFN-Sezione di Catania, Via S. Sofia 64, 95127 Catania, Italy}

\author{Paolo Castorina}

\affiliation{INFN-Sezione di Catania, Via S. Sofia 64, 95127 Catania, Italy}
\affiliation{Institute of Particle and Nuclear Physics, Faculty of Mathematics and Physics,
Charles University, V  Holešovičkách 2, 18000 Praha 8, Czech Republic}

\author{Dario Zappalà}

\affiliation{INFN-Sezione di Catania, Via S. Sofia 64, 95127 Catania, Italy}

\author {Luigi Amico}

\thanks{On leave from Dipartimento di Fisica e Astronomia ``Ettore Majorana'', University of Catania.}

\affiliation{Quantum Research Centre, Technology Innovation Institute, Abu Dhabi, UAE}

\affiliation{INFN-Sezione di Catania, Via S. Sofia 64, 95127 Catania, Italy}

\affiliation{Centre for Quantum Technologies, National University of Singapore, 3 Science Drive 2, Singapore 117543, Singapore}

\maketitle

\date{\today}

{\textit{Introduction} --} One of the most striking features of  quantum chromodynamics (QCD) is  the quarks' confinement leading to hadrons\cite{di1996mechanisms}. 
 Even though important insights into the problem have been achieved, confinement physics remains difficult to explore. The standard path that has been widely followed in the literature is to restrict the ambient space of the theory \cite{schwinger1962gauge,coleman1976more} and/or by coarse graining the space time\cite{wilson1974confinement}. 
The resulting theories are  known as lattice gauge theories\cite{kogut1983lattice,kogut1979introduction}. Besides benchmarking and regularising continuous gauge theories, lattice gauge theories have defined an entirely new direction of condensed matter
\cite{wegner2015duality,wen2004quantum,wen2017colloquium,broholm2020quantum} and  quantum information $\&$ computation\cite{kitaev2003fault,kitaev2006anyons,kitaev2010topological,bravyi2002fermionic}.

In the era of quantum technologies, several schemes and physical platforms  simulating  lattice gauge theories have been proposed, ranging from superconducting circuits \cite{mildenberger2022probing,marcos2014two,irmejs2022quantum,roy2023soliton} and ion traps\cite{davoudi2020towards} to cold atoms\cite{aidelsburger2022cold,schweizer2019floquet,mil2020scalable,surace2020lattice,surace2023ab,de2018accurate,bernien2017probing,yang2020observation,zhao2022thermalization,halimeh2022tuning}.
Indeed, insights in  important and challenging aspects in the field   have  been explored\cite{dalmonte2016lattice,banuls2020simulating,shaw2020quantum,gonzalez2020rotor}. 
In this context, analogic quantum simulations that explore finite baryonic densities and real time dynamics are of particular relevance \cite{nagata2022finite,montangero2016real,martinez2016real}.

\begin{figure}
\includegraphics[width=\linewidth]{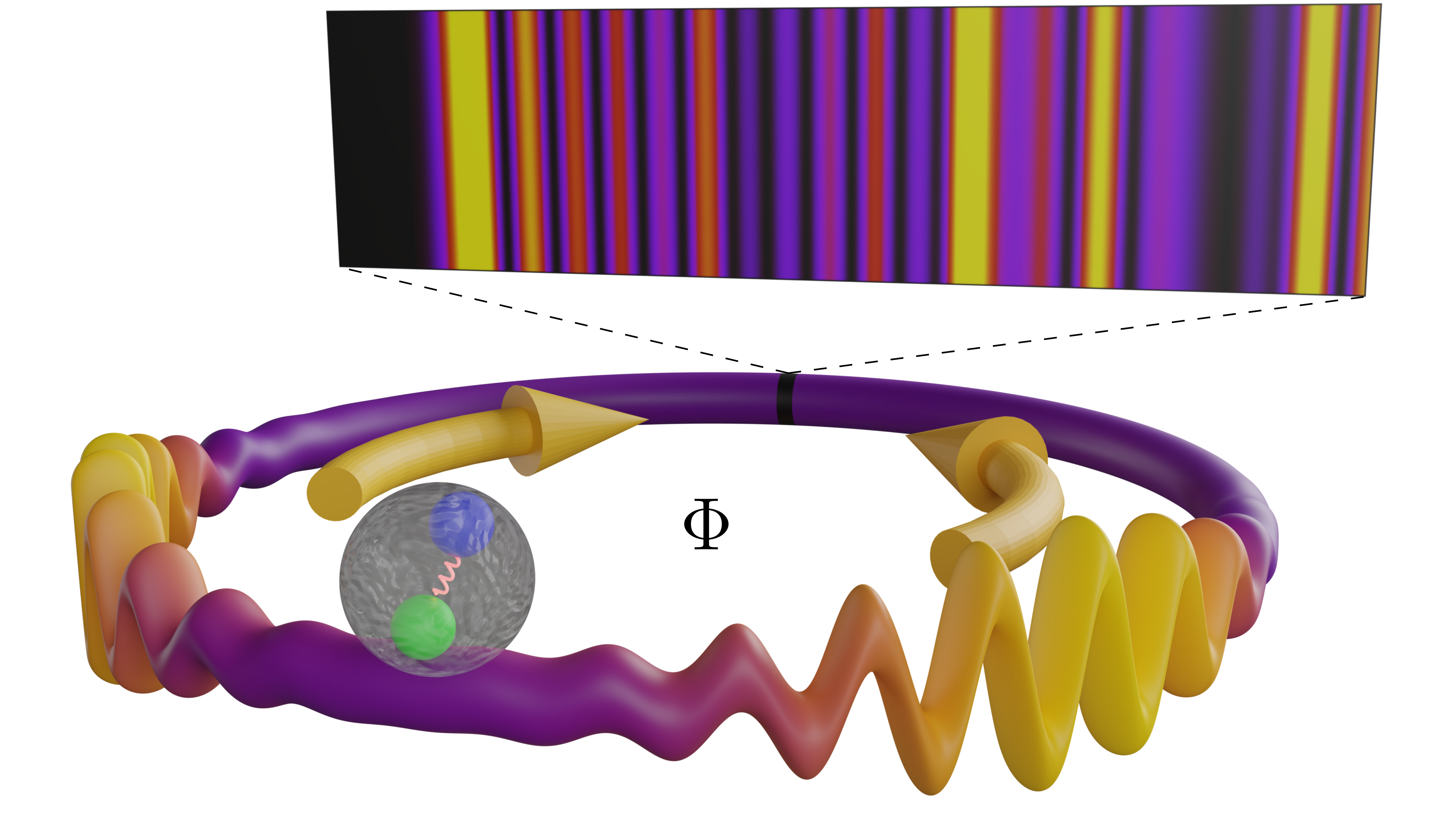}
\caption{\textit{Lattice gauge theory on a ring of mesoscopic scale}. Illustration of meson dynamics on a ring pierced by a magnetic flux $\Phi$. The screen captures the Aharonov-Bohm oscillations produced by clockwise and anti-clockwise components of the wave function.}
\label{ring}
\end{figure}

In this paper,  we analyse  the real time dynamics of  gauge theories at mesoscopic spatial scales. Indeed, mesoscopic physics has been providing  an important avenue to  highlight fundamental notions in quantum mechanics such as macroscopic quantum superposition, persistent currents, coherent quantum dynamics, etc\cite{imry2002introduction}. With a similar logic, here we demonstrate how, through mesoscopic platforms, new coherent properties of the gauge theories can be disclosed. Specifically, we consider a gauge theory with non-trivial confinement properties, and constrain its dynamics to a ring circuit pierced by an effective magnetic field - see Fig.\ref{ring}. Such configuration is well within the experimental capabilities of quantum technologies \cite{amico2022colloquium}. Here we provide a specific scheme implementing the theory through driven  networks of Rydberg atoms.   In such  conditions, we will demonstrate that confined matter experiences a Aharonov-Bohm effect with unique features,  disclosing fundamental features of the confinement physics of the underlying  lattice gauge theories \cite{imry2002introduction,buttiker1985generalized,imry1999conductance,gefen1984quantum}.  For the quantitative analysis, we consider  a $Z_2$ gauge theory whose confined matter can define the analog of QCD mesons with a confining potential that linearly increases with the quarks separation \cite{borla2020confined,borla2021gauging,surace2021scattering,kormos2017real}.
We refer to the minimal, but interesting setting of the analog of mesons made of a  single quark pair\cite{di1996mechanisms}. As a remarkable  aspect of such system,  we note that confined 'quark' pair's relative coordinate and center of mass result to be tightly coupled.  Such a feature implies a specific Aharonov-Bohm effect and  dramatic  effects in the meson current dynamics. Such phenomena can characterize the dynamics of the internal structure of the confined matter at a length scale that would be very hard, if not impossible, to explore with standard approaches in  particle and  accelerators physics.

{\textit{Methods} --}
The model under consideration is a one-dimensional $\mathbb{Z}_2$-lattice gauge theory of spinless fermions, hopping in a ring-shaped lattice and pierced by an effective magnetic flux $\Phi$. Such a  theory shows the confinement of   fermions into mesons\cite{borla2020confined,kebric2021confinement,surace2021scattering}. The Hamiltonian reads
\begin{equation}
\label{Hamiltonian}
    \mathcal{H} = \sum_{j=1}^L w \, (e^{i \frac{2\pi}{L} \Phi/\Phi_0} c_j^\dagger c_{j+1} + h.c.) \, \sigma_{j+\frac{1}{2}}^x + \frac{\tau}{2} \sigma_{j+\frac{1}{2}}^z \, .
\end{equation}
The fermionic matter $c_j$ lives in the sites of the lattice. Gauge fields, represented by spin $1/2$ variables residing on the bonds connecting nearest neighbor  sites, act as a $\mathbb{Z}_2$ propagator:  when particles hop across neighbor sites, a  spin-flip occurs in the corresponding bond. The string tension $\tau$ assigns quantum dynamics to the gauge field.   Such tension is responsible for the confinement of fermions into mesons:  $\tau=0$ corresponds to a deconfined phase; for finite tension, specific  bound states can be formed with an energy that is increasing with $\tau$. The theory \eqref{Hamiltonian} is invariant under global $U(1)$ transformations of the fermionic operators $c_j \to e^{i \alpha} c_j$ and under local gauge transformations generated by the operators $G_j = \sigma_{j-\frac{1}{2}}^z (-1)^{n_j} \sigma_{j+\frac{1}{2}}^z$. Since $\commutator{H}{G_j} = 0 \, \forall j$, the dynamics is restricted to gauge sectors, singled out by  the eigenvalues of $G_j$:  $G_j|\psi \rangle=\pm |\psi \rangle \, 
 \forall j$ (Gauss law). 

 Periodic boundary conditions constraint the fermion parity to be even or odd, depending on the number of negative eigenvalues of $G_j$ along the lattice, as $P = \prod_j (-1)^{n_j} = \prod_j G_j$. Here, we will work in the neutral gauge sector $G_j = 1 \, \forall j$.

To address the matter current, we refer to  the current  $\displaystyle{\mathcal{I} = - {\partial \langle \mathcal{H}  (\Phi)\rangle}/{\partial \Phi}}$, which is related to the $U(1)$ symmetry of the model \eqref{Hamiltonian}; we note that $\commutator{\mathcal{H}}{\mathcal{I}} \neq 0$ only if $\tau \neq 0$. As we shall see, the current can track relevant features of  the confinement-deconfinement transition. 

{\textit{Floquet implementation of the $\mathbb{Z}_2$-lattice gauge theory on a ring pierced by a synthetic field} --} Here, we demonstrate how the the dynamics entailed by the Hamiltonian (\ref{Hamiltonian})  can effectively be obtained by Floquet engineering. We first  note that, by fixing the  gauge as  $G_j = 1 \, \forall j$, we can eliminate matter degrees of freedom by $n_j = \frac{1}{2} \left(1 - \sigma_{j-\frac{1}{2}}^z \sigma_{j+\frac{1}{2}}^z \right)$. By employing the mapping $\sigma_{j+\frac{1}{2}}^z \to \sigma_j^z$ and $(c_j^\dagger - c_j) \sigma_{j+\frac{1}{2}}^{x/y} (c_{j+1}^\dagger + c_{j+1}) \to \sigma_{j}^{x/y}$, the Hamiltonian \eqref{Hamiltonian} can be identified as the  Ising-type model
\begin{align}
\nonumber
    \mathcal{H}_{\text{spin}} = &\sum_{j=1}^L w\Biggl[ \cos(\frac{2\pi\Phi}{L\Phi_0}) \frac{(1-\sigma_{j-1}^z \sigma_{j+1}^z)}{2} \, \sigma_j^x \\
\label{eq:spin_model}
    &+ \sin(\frac{2\pi\Phi}{L\Phi_0}) \frac{(\sigma_{j-1}^z - \sigma_{j+1}^z)}{2} \, \sigma_j^y \Biggr] + \frac{\tau}{2} \sigma_j^z \, .
\end{align}
\cite{surace2021scattering,borla2021gauging}.
Indeed, the Hamiltonian (\ref{eq:spin_model}) can be obtained  as lowest order of the high frequency Floquet driving of  $\mathcal{H}_0 = \sum_j \big[ \frac{\lambda}{2} \, \sigma_j^x + \frac{\tau}{2} \, \sigma_j^z + (-1)^j \, J \, \sigma_j^z \sigma_{j+1}^z \big]$ 
with  $\mathcal{H}_{\text{drive}}(t) = \sum_j \frac{A}{2} \cos(\Omega t - (-1)^j \varphi) \sigma_j^z$, in which  $J \sim \Omega$, $\varphi = \frac{\pi \Phi}{2L \Phi_0}$, $w = \frac{\lambda}{2} \mathcal{J}_4\left(\frac{A}{\Omega}\right)$, $\mathcal{J}_4(x)$ being the 4-th order Bessel function of the first kind; $A/\Omega$ is chosen as a zero of $\mathcal{J}_0(x)$ - see Appendix \ref{app:floquet_section} for the analytical details. 
Such considerations, grant our theory  to be  implemented  through a ring-shape network of Rydberg atoms trapped in a optical tweezers \cite{leseleuc2018accurate,browaeys2020many}; the Floquet protocol can be achieved by a suitable local driving of the atoms detuning - see Appendix \ref{app:floquet_section}.

{\textit{Current of mesons in the ground state} --} 
For $\tau = 0$, fermions are deconfined. When subjected to an external magnetic flux, free fermions are known to display saw-tooth oscillations in the current, with period of  $\Phi_0$\cite{leggett1991}. In contrast, here we observe a saw-tooth behavior with $\Phi_0/2$ (see also Appendix \ref{app:ground_current}).  Such  effect is  due to the gauge field degrees of freedom. For $\tau=0$, the Hamiltonian  is indeed block diagonal in the basis of the eigenfunctions of the Wilson loop operator $W = \prod_j \sigma_{j+\frac{1}{2}}^x$, with eigenvalues $\pm 1$. The fractionalization of the current is thus produced as an effect of the level crossings between the ground states of the two blocks - see Appendix \ref{app:ground_current}. 
In the confined regime $\tau \neq 0$,  the Hamiltonian is not block diagonal anymore. Nevertheless, the  ground state current is still found with a halved periodicity because of the formation of confined bound states, in which the participating two particles share a single flux-quantum. In the limit of large $\tau/w$, this analysis is supported by a second order perturbative expansion - see Appendix \ref{app:ground_current}.

Remarkably, in the two-particles sector, we find that the maximum of the ground state current ${\cal I}_{max}$  displays scaling behaviour as function of the string tension and of the ring size $L$ - see Appendix \ref{app:ground_current}. For large enough $\tau$, the scaling function provides the particle-like behaviour \cite{imry2002introduction} $\mathcal{I}_{max} \sim L^{-2}$, from which it deviates substantially in the regime of small $\tau$.
The interplay between $\tau$ and $L$ emerges also in the correlation function $g(r) = \langle c^\dagger_{j+r} \prod_{l = j}^{j+r-1} \sigma^x_{l+\frac{1}{2}} c_j \rangle$ \cite{borla2020confined,kebric2021confinement}. The gauge invariant correlator $g(r)$ displays oscillations with $r$ which are reminiscent of a free particle behavior; these oscillations are exponentially damped, with the law $g(r) \sim e^{-r/\xi}$, in which $\xi\sim \tau^{-\gamma(L)}$, with   $\gamma$ displaying a weak dependence on $L$, being $\gamma \sim 0.52$ for large enough $L$ - see Appendix \ref{app:correlator_section}.

{\textit{A single meson in a ring pierced by an effective magnetic field} --} 
 The single meson on the ring is described by the state  
$ 
    \ket{\psi} = \sum_{j_1,j_2;\sigma} \psi^\sigma(j_1,j_2) c^\dagger_{j_1}c^\dagger_{j_2} \ket{0,\sigma} \, 
$. We take $\sigma$ as the eigenvalue of $-\sigma^z_{L+\frac{1}{2}}$, distinguishing  the two allowed spin configurations corresponding to fixed particles positions $j_1 < j_2$. We solve the first quantized spectral problem corresponding to Eq.(\ref{Hamiltonian}) exactly:
$
\mathcal{H}^\sigma \psi^\sigma(j_1,j_2)=E\psi^\sigma(j_1,j_2)  
$

\begin{eqnarray}
\label{eigenfunction}
&&\psi^\sigma_{K,l}(s,r)=e^{iKs} \phi_{K,l}^\sigma(r) \quad, \; K = \frac{2\pi n}{L}  \\
&&    \phi_{K,l}^\uparrow(r) = \mathcal{N} \left\{ \frac{\mathcal{J}_{\frac{E_{K,l}}{\tau} - r }\left(\frac{2 \, w_K}{\tau}\right)}{\mathcal{J}_{\frac{E_{K,l}}{\tau}}\left(\frac{2 \, w_K}{\tau}\right)} - \frac{\mathcal{Y}_{\frac{E_{K,l}}{\tau} -r}\left(\frac{2 \, w_K}{\tau}\right)}{\mathcal{Y}_{\frac{E_{K,l}}{\tau}}\left(\frac{2 \, w_K}{\tau}\right)} \right\} \, , \nonumber 
\end{eqnarray}
with $\psi_{K,l}^\downarrow(s,r)$ obtained as $-\psi_{K,l}^\uparrow(s+L/2,L-r)$. The variables $s = (j_1 + j_2)/2$ and $K$ are the centre of mass coordinate and momentum respectively, while $r = j_2 - j_1$ is the  relative coordinate. For future reference, we define the string length as $R = r$, if $\sigma = \uparrow$, or $R = L-r$, if $\sigma = \downarrow$: it represents the length of the string of spin ups comprised between the two fermions forming a meson. $\mathcal{J}$ and $\mathcal{Y}$ are Bessel functions of the first and of the second kind respectively,  whose order is fixed by the energies and particle relative positions \cite{lagnese2020confinement}; their argument figures $w_K(\Phi) = 2 \,w \, cos\left(\frac{K}{2} + \frac{2 \pi \Phi}{L \Phi_0} \right)$ providing the centre of mass energies. The energy eigenvalues are obtained  by requiring that  $\phi_{K,l}(L) = \phi_{K,l}(0)=0$, corresponding to the Pauli constraint on periodic boundary conditions; as such, energies are labelled by an index $l$ and depend, through $w_K$,  on $K$ and $\Phi$: $E=E_{K,l}(\Phi)$.  
As Eq.\eqref{eigenfunction} shows, $r$ and $s$ are indeed coupled: the motion of the relative coordinate  can be obtained after the center of mass dynamics is specified in both the eigenfunctions and energies. As a result, the magnetic field couples to {\it both meson center of mass and relative coordinates}.
Such a remarkable effect is originated by the lattice,  that in the present case is a defining feature of the theory.

In this formalism, general features of the meson dynamics can be grasped. Specifically, the Heisenberg equations of motion can be integrated exactly. 
The string length is seen to perform characteristic Bloch oscillations \cite{collura2022time}. Due to the coupling between $s(t)$ and $R(t)$, we find that the Bloch oscillations of the string length propagate to oscillations of the center of mass position, with a velocity which is fixed by ${\cal I}$ - see Appendix \ref{app:twobodydynamics}.

\begin{figure}[th!]
    \centering
    \includegraphics[width = 0.95 \linewidth]{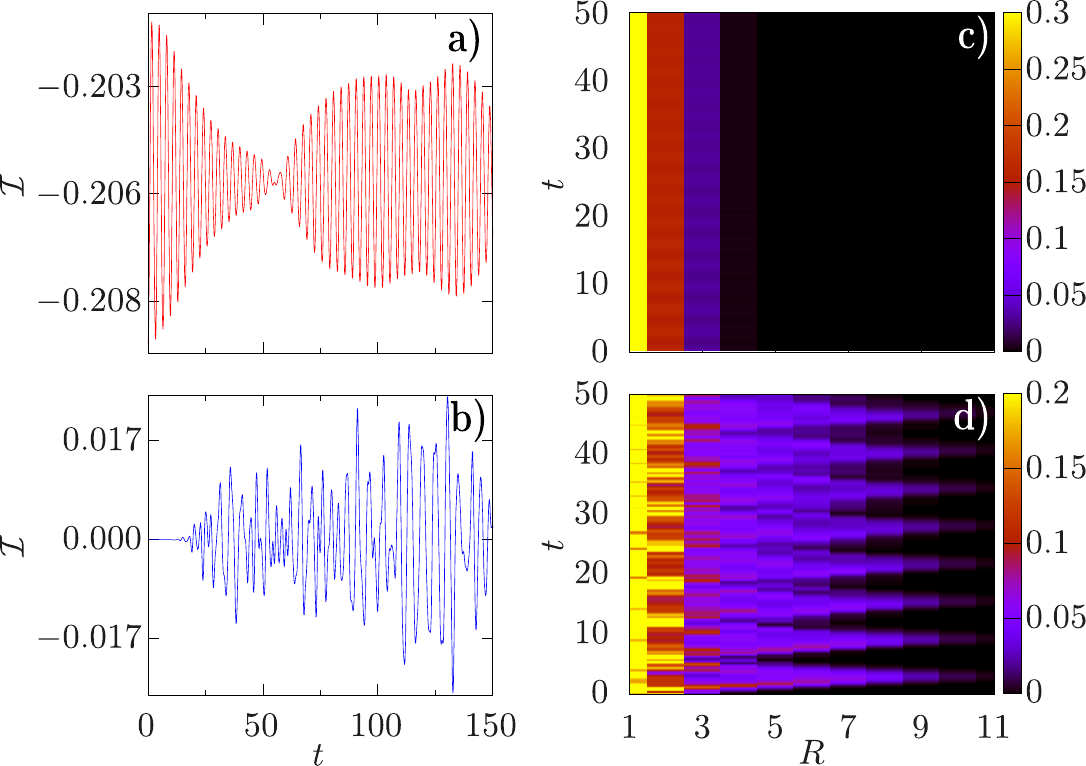}
       
    \caption{A single meson whose center of mass has been localized within a gaussian distribution of width $\Sigma$, as in Eq.(\ref{eqn:initialstate}), is quenched by applying the Hamiltonian (\ref{Hamiltonian}) as in Eq.\eqref{eq:quench}. To this end, the flux is quenched to the value $\Phi/\Phi_0 = 0.8$. The average current (representing the center of mass velocity) and the probability to find the meson with string length $R$ (labelling the length of a string of spin ups connecting the two bound particles), $\mathcal{I}(t)$ and $P(R) = \sum_s |\braket{s,R}{\psi(t)}|^2$, are displayed in panels a),b) and c),d) respectively. The upper panels refer to $\Sigma = 2$, $\tau = 1$; the lower panels  refer to $\Sigma = 10^{-6}$, $\tau = 1$. In all cases, we consider $L = 21$ sites.}
    \label{fig:quench_dynamics}
\end{figure}

{\textit{Quench dynamics} --}
Here, we study the effects of the aforementioned coupling between center of mass and relative coordinates degrees of freedom - see Eq.\eqref{eigenfunction} - that emerge in the dynamics of a single meson. To this end, we apply a dynamical  protocol in which a single meson, whose center of mass is localized at $s=s_0$, is put in motion by quenching the effective magnetic field:  
\begin{equation}
\label{eq:quench}
\psi^\sigma_\Phi(t)=e^{-i H(\Phi) t} \psi^\sigma_0(t=0) \, .
\end{equation}

\ni The protocol is initialized with  $\psi_0(t=0)$ describing a meson 
with center of mass $s$ localized within a Gaussian distribution:
 \begin{equation}
 \label{eqn:initialstate}
 \psi_0^\sigma(t=0)= e^{-{{(s + L/2 \, \delta_{\sigma,\downarrow}-s_0)^2}/(2\, \Sigma)}} \, \psi^\sigma_{K=0,l=1}(s,r) \, .
 \end{equation}
 Such a state  excites center of mass momenta within values that depend on the Gaussian variance $\Sigma$.
 This way, the dynamics can effectively explore different portions  of the meson spectrum. As figures of merit, we consider the average value of the current $\bra{\psi(t)}\mathcal{I}\ket{\psi(t)}$ and the string dynamics, which we address by calculating $P(R,t) = \sum_{s} \abs{\braket{s,R}{\psi(t)}}^2$. In the deconfined regime $\tau=0$, the current is a constant of the motion. For $\tau \neq 0$, a distinctive oscillatory dynamics arises, depending on the width of the initial gaussian localization. For finite values of $\Sigma$, the dynamics  starts as soon as the quench is applied and the current reflects a directional center of mass motion - see Fig.\ref{fig:quench_dynamics} a).
In the limit of a sharply localized (delta-like) center of mass, we observe that $\mathcal{I}$
remains nearly vanishing for a characteristic time $T(\tau)$  increasing algebraically with $\tau$, reaching the linear dependence   for large confinement - see  Fig.\ref{fig:quench_dynamics} b). 
For $t > T(\tau)$, the current starts oscillating around zero, at a faster pace with increasing $\tau$.
The observed oscillatory behavior reflects interference effects between energy levels with different quantum numbers $l$ and reflects the oscillations of the string length - see Appendix \ref{app:twobodydynamics}. By increasing the string tension, fewer and fewer levels are involved in the dynamics, giving rise to clear beating modes.
 
 The string is seen to perform periodic oscillations whose amplitude and period depend on the gaussian width $\Sigma$ and on the value of the string tension $\tau$- see Fig.\ref{fig:quench_dynamics} c),d). We find that $\Sigma$ and $\tau$ have similar effects: to smaller values of $\Sigma$, $\tau$ correspond slower and broader oscillations. Indeed, in the limit of a delta-like center of mass localization, all center of mass momenta are equally populated, thus exciting larger portions of the meson spectrum. Similarly, for small $\tau$, an increasing number of energy levels are involved in the dynamics, as the particles are more free to move.

\begin{figure}[th!]
    \centering
    \includegraphics[width = 0.95 \linewidth]{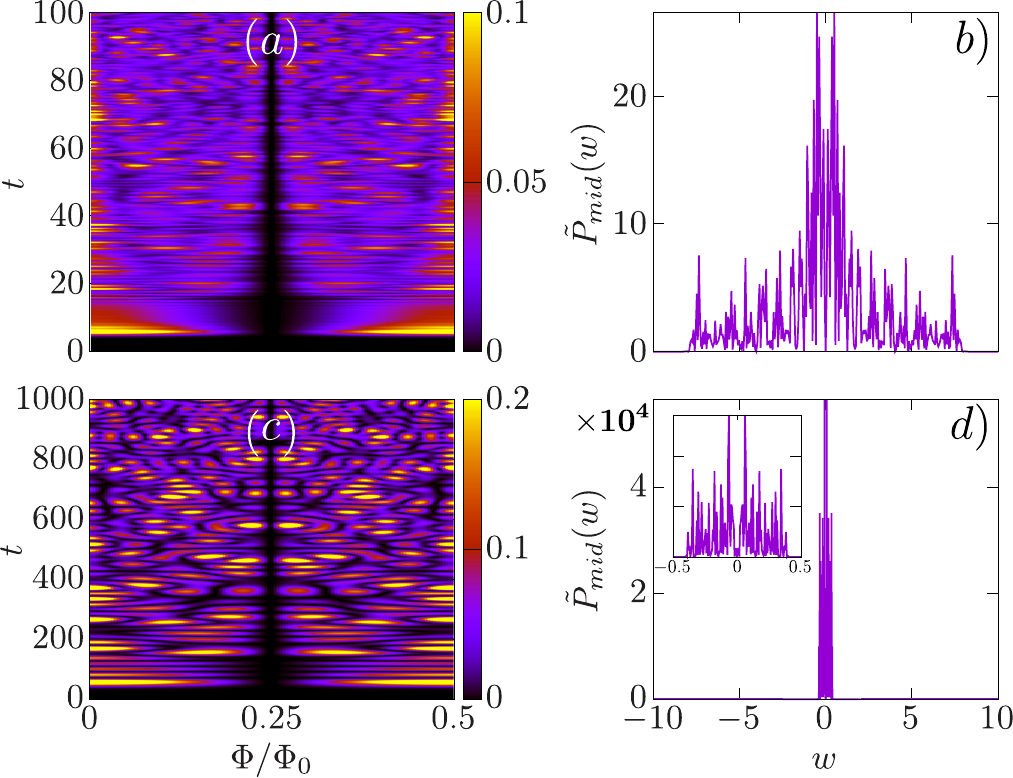}
    \caption{\textit{Analysis of the meson Aharonov-Bohm oscillations}. A single meson wave function, whose center of mass is initialized in the site $s_0$, as specified in Eq.(\ref{eqn:initialstate}), evolves along the ring and is monitored at $s=s_0 + L/2$. Here $L = 20$. The upper and the lower panels are respectively for $\tau = 0.1$ and $\tau = 10$. The panels (a),(c) display $P_{mid}(t) = \sum_r |\braket{s_0 + L/2,r}{\psi(t)}|^2 $. Panels (b),(d) display the Fourier transform of $\left(P_{mid}(t) - \overline{P_{mid}}\right) = \tilde{P}_{mid}(w)$ at $\Phi/\Phi_0 = 0$, where $\overline{P_{mid}}$ is the time average of $P_{mid}(t)$. The figure in panel (c) is plot for $t$ up to $1000$: indeed, for large $\tau$, the relevant time scale is set by $t/\tau$ - see also text.}
    \label{fig:aharonov}
\end{figure}

{\textit{Meson Aharonov-Bohm oscillations} --}
Here, we study the  Aharonov-Bohm effect experienced by a meson as function of its confinement, within the same quench protocol that is described in the previous paragraph. After the quench, the wave function evolves along the ring, with
clockwise and anti-clockwise probability waves that can eventually produce interference - see Fig.\ref{ring}.
We will see how the interplay between confinement and the imparted phase $\Phi$ establishes a peculiar dynamics of the meson, with characteristic Aharonov-Bohm oscillations. For a quantitative analysis, we monitor the probability density at  $s= s_0 + L/2$ - see Fig.\ref{ring}. 
The interference fringes display a marked dependence on the string tension whose increase progressively localizes the meson - see Fig.\ref{fig:aharonov}.
As a first feature, we note that the interference fringes are generically blurred by decreasing $\tau$. Such a feature points to a reduction of the number of distinct modes characterizing the wave functions as the string tension is increased (from  $k=2\pi/L$ for strongly confined meson, to a nearly double $k$ number for loosely paired particles). Because of the aforementioned coupling between $s$ and $R$,  the response of the meson to the magnetic field is highly non trivial: while, for strong confinement, the meson is characterized by a single particle Aharonov-Bohm effect, for weaker confinements Aharonov-Bohm and relative coordinate dynamics combine together to produce distinctive Aharonov-Bohm oscillations. This feature is reflected in the time dynamics of the probability density, which is quantitatively analysed via Fourier transform. The number of the involved frequencies is strongly reduced at large values of $\tau$, reflecting a suppression of the relative degrees of freedom, in favour of an emerging one-particle picture (see Appendix \ref{app:ab_section} for the single particle behaviour). The distribution of such frequencies around zero is significantly broader for smaller values of $\tau$. In this regime, the dynamics excites energy levels $E_{K,l}$ ranging over several values of $l$. On the other hand, larger values of $\tau$ correspond to larger gaps among energy levels with different $l$, effectively constraining the dynamics only within the lowest relative levels. We note that larger values of $\tau$ correspond to slower dynamics, whose time scale is fixed by $t/\tau$ for $\tau \gg 1$ - see Appendix \ref{app:ab_section} for analytical details. This analysis is sensitive to the value of the applied external flux, manifesting flux-driven interference effects. For $\Phi/\Phi_0=1/4$,  a completely destructive interference at $s = s_0 + L/2$ results. This interference reflects the meson nature as two particle bound state. As we discuss below, such a feature emerges also in the meson current. 

{\textit{Conclusions}--}
 In this work, we studied the mesoscopic coherence inherent in lattice gauge theories. To this end, we made the theory live on a  ring-shaped lattice  and analysed its  response to the  magnetic field.  
 As paradigmatic case, we considered a $\mathbb{Z}_2$ lattice gauge theory providing meson confinement.
The magnetic field  couples both to the confined matter center of mass and internal structure. We demonstrate how such theories can be analogically implemented through Floquet-driven networks of Rydberg atoms (see also Appendix \ref{app:floquet_section}).

We find a dynamics and interference process that are beyond the particle-like phenomena and reflecting the meson confinement. In particular, a peculiar kind of Aharonov-Bohm effect occurs that reveals key features of the confinement phenomenon in its own: while a tightly confined meson displays point-like particles Aharonov-Bohm effect (with oscillations resulting to be dependent only on $\Phi$), for weaker confinements Aharonov-Bohm oscillations display clear signature of an entangled combination of the meson internal (oscillatory) dynamics and the meson's motion as a whole - see Fig. \ref{fig:aharonov} and Appendix \ref{app:ab_section}.         
 Relevant insights in the coherence of meson dynamics is obtained by studying the mesoscopic current. 
In the ground state, the  current  periodicity with the effective magnetic field reflects the feature of  mesons bound states and singles out the contribution of the gauge field\cite{leggett1991} - see Appendix \ref{app:ground_current}.
The scaling of the meson current that we found, involving a suitable combination of $\tau$ and $L$,  addresses the mesoscopic coherence of the  meson in the ground state quantitatively - see Appendix \ref{app:ground_current}. Aharonov-Bohm oscillations and mesoscopic coherence combine together in the current quench dynamics - see Fig. \ref{fig:quench_dynamics}. On one hand, 
the slow and fast  frequency current dynamics reflects the interplay between  the meson center of mass and the relative coordinate motion.
On the other hand, the current assumes non vanishing values after a characteristic time depending on the confinement and on the meson localization.
In this time, clock-wise and anticlock-wise moving components of the meson wave function accumulate a topological phase that,  depending on the value of $\Phi$, can start the matterwave  current. 

Exploiting the state of the art of quantum technology, our predictions can be tested through  the Rydberg atoms system we detailed in the present paper (see also Appendix \ref{app:floquet_section}), and possibly also  in cold atoms platforms\cite{schweizer2019floquet,mil2020scalable,surace2020lattice,surace2023ab,de2018accurate,bernien2017probing,yang2020observation,amico2022colloquium}, or  through driven superconducting circuits\cite{mildenberger2022probing,marcos2014two,irmejs2022quantum,roy2023soliton}. This way, by studying the real time out of equilibrium coherence dynamics of lattice gauge theories can define  a new framework for the cross-fertilization between high-energy physics and quantum technology. 

One interesting avenue for future research would be to extend the present analysis to larger gauge groups, such as $\mathbb{Z}_N$ \cite{rigobello2021entanglement}. Confinement can be further analysed by exploiting alkali-earth SU$(N)$ fermions\cite{cazalilla2014ultracold,rapp2007color,werner2022spectroscopic}.

\begin{acknowledgments}
{\textit{Acknowledgments} --} We thank Alejandro Bermudez, Francesco Perciavalle, Gianluigi Catelani, Juan Polo and Wayne J. Chetcuti for discussions.
\end{acknowledgments}

\begin{appendix}
    \section{Floquet implementation of the $\mathbb{Z}_2$-lattice gauge theory on a ring pierced by a synthetic field}
    \label{app:floquet_section}
Here we show how we can implement the Hamiltonian 
\begin{equation}
\label{eq:app:model}
    \mathcal{H} = \sum_j \Bigl[ w \left(e^{\frac{2 i \pi \Phi}{L \Phi_0}} c_j^\dagger \, \sigma^x_{j+\frac{1}{2}} \, c_{j+1} + h.c. \right) + \frac{\tau}{2} \, \sigma^z_{j + \frac{1}{2}} \Bigr] \, .
\end{equation}

\ni when explicitly restricted to the neutral gauge sector and in the presence of a synthetic flux $\Phi$. 

In the neutral gauge sector, the explicit resolution of the gauge constraints $G_j = \sigma_{j-\frac{1}{2}}^z (-1)^{n_j} \sigma_{j+\frac{1}{2}}^z = 1 \, \forall j$, allows for the elimination of the fermionic degrees of freedom, since $n_j$ can be rewritten as 
\begin{equation}
\label{app:eq:domains}
    n_j = \frac{1}{2} \left(1- \sigma_{j-\frac{1}{2}}^z \sigma_{j+\frac{1}{2}}^z\right)
\end{equation}

\ni By exploiting Eq.\eqref{app:eq:domains} and the mapping $\sigma_{j+\frac{1}{2}}^z \to \sigma_{j}^z$, $(c_j^\dagger 
- c_j) \sigma_{j+\frac{1}{2}}^{x/y} (c_{j+1}^\dagger + c_{j+1}) \to \sigma_j^{x/y}$ \cite{surace2021scattering,borla2021gauging}, the system can be identified with the following Ising-type model:
\begin{align}
\label{app:eq:hspin}
\nonumber
    \mathcal{H}_{\text{spin}} &= \sum_{j=1}^L w \Biggl[ \cos(\frac{2\pi\Phi}{L\Phi_0}) \frac{(1-\sigma_{j-1}^z \sigma_{j+1}^z)}{2} \, \sigma_j^x \\
    &+ \sin(\frac{2\pi\Phi}{L\Phi_0}) \frac{(\sigma_{j-1}^z - \sigma_{j+1}^z)}{2} \, \sigma_j^y \Biggr] + \frac{\tau}{2} \sigma_j^z \, .
\end{align}

\ni In this picture, fermions are mapped into domain walls separating domains of reversed magnetization in a spin chain, as is explicit from Eq.\eqref{app:eq:domains}. The hopping of fermions is then mapped into hopping of domain walls. Indeed, the square-bracketed terms in Eq.\eqref{app:eq:hspin} produce a spin flip at site $j$, under the condition that the $\sigma^z$ variables at the neighboring sites $j\pm1$ have opposite direction, as enforced by the constraints $\frac{(1-\sigma_{j-1}^z \sigma_{j+1}^z)}{2}$ and $\frac{(\sigma_{j-1}^z - \sigma_{j+1}^z)}{2}$. This produces a hopping of the domain wall that separates the spins at either positions $j-1$ and $j$, or $j$ and $j+1$. The term $\frac{(\sigma_{j-1}^z - \sigma_{j+1}^z)}{2} \, \sigma_j^y$ provides directionality, as it distinguishes leftwards from rightwards hopping processes.

We propose an implementation of Eq.\eqref{app:eq:hspin}, emerging as the effective description of a periodically driven hamiltonian $\mathcal{H}(t) = \mathcal{H}_0 + \mathcal{H}_{\text{drive}}(t)$ with driving frequency $\Omega$, at the lowest order in the large $\Omega$ limit. In particular, $\mathcal{H}_0$ is taken as an Ising model in the presence of longitudinal and transverse fields
\begin{equation}
\label{app:eq:ising}
    \mathcal{H}_0 = \sum_j \left[\frac{\lambda}{2} \, \sigma_j^x + \frac{\tau}{2} \, \sigma_j^z + (-1)^j \, J \, \sigma_{j}^z \sigma_{j+1}^z \right] \, ,
\end{equation}

\ni where periodic boundary conditions are assumed. We choose to drive the longitudinal field, according to 
\begin{equation}
\label{app:eq:drive}
    \mathcal{H}_{\text{drive}}(t) = \sum_j \frac{A}{2} \cos(\Omega t - (-1)^j \varphi) \, \sigma_j^z
\end{equation}

The model \eqref{app:eq:ising} can be naturally implemented on an array of Rydberg atoms trapped in optical tweezers , where the transverse and longitudinal fields would respectively correspond to the Rabi frequency and to the detuning terms that are induced by laser-driving the ground to Rydberg-state transition of the atoms, in the rotated frame \cite{leseleuc2018accurate,browaeys2020many}.
The scheme that we propose would then correspond to a local periodic modulation of the atoms detuning - see Eq.\eqref{app:eq:drive}.

We notice that the transverse field term in Eq.\eqref{app:eq:ising} can be rewritten as $\sum_j w \left[ \frac{(1-\sigma_{j-1}^z \sigma_{j+1}^z)}{2} \, \sigma_j^x + \frac{(1+\sigma_{j-1}^z \sigma_{j+1}^z)}{2} \, \sigma_j^x\right]$: spin flips caused by $\sigma_j^x$ can either move domain walls (first term) or create/annihilate them in neighboring pairs (second term). We work under the assumption that $J \sim \Omega$. The large staggered Ising interaction results in a suppression of the hopping of domain walls. We will show that the hopping is restored in the high-frequency expansion, assisted by an additional "Peierls phase". Additionally, the terms that create/annihilate domain walls in pairs can be suppressed by an appropriate choice of the driving amplitude.

We perform the time-dependent unitary transformation $\mathcal{H}'(t) = U^\dagger(t) \mathcal{H}(t) U(t) - i U^\dagger(t) \partial_t U(t)$, with
\begin{equation}
    U(t) = \exp{-i \sum_j \left[ \frac{\chi_j(t)}{2} \, \sigma_j^z + (-1)^j \Omega t \, \sigma_j^z \sigma_{j+1}^z \right]}
\end{equation}
\ni and $\chi_j(t) = \frac{A}{\Omega} \sin(\Omega t - (-1)^j \varphi)$. It leads to:
\begin{widetext}
\begin{align*}
     &\mathcal{H}'(t) = \sum_j \Bigl\{ \frac{\tau}{2} \, \sigma_j^z + (-1)^j (J-\Omega) \, \sigma_j^z \sigma_{j+1}^z + \frac{\lambda}{4} \bigl[ (1+\sigma_{j-1}^z \sigma_{j+1}^z) (\cos(\chi_j(t)) \, \sigma_j^x - \sin(\chi_j(t)) \, \sigma_j^y) \\ 
    & + \cos(4 \Omega t) (1 - \sigma_{j-1}^z \sigma_{j+1}^z) (\cos(\chi_j(t)) \, \sigma_j^x - \sin(\chi_j(t)) \, \sigma_j^y) + (-1)^j \sin(4\Omega t) (\sigma_{j-1}^z - \sigma_{j+1}^z) (\cos(\chi_j(t)) \, \sigma_j^y + \sin(\chi_j(t)) \, \sigma_j^x)\bigr] \Bigr\} \, .
\end{align*}
\end{widetext}

\ni The stroboscopic dynamics described by $\mathcal{H}'(t)$, observed at multiples of the driving period $T=\frac{\Omega}{2\pi}$ in the limit of large $\Omega$, is effectively described by the first order, time independent Floquet Hamiltonian $\mathcal{H}_{\text{eff}} = \frac{1}{T} \int_0^T dt \, \mathcal{H}'(t)$. Under the assumption that $J \sim \Omega$, this results into
\begin{widetext}
\begin{equation*}
    \mathcal{H}_{\text{eff}} = \sum_j \left\{ \frac{\tau}{2} \, \sigma_j^z + \frac{\lambda}{4} \mathcal{J}_0\left(\frac{A}{\Omega}\right) (1 + \sigma_{j-1}^z \sigma_{j+1}^z) \, \sigma_j^x + \frac{\lambda}{4} \mathcal{J}_4\left(\frac{A}{\Omega}\right) \left[\cos(4 \varphi)(1-\sigma_{j-1}^z \sigma_{j+1}^z) \, \sigma_j^x + \sin(4 \varphi)(\sigma_{j-1}^z - \sigma_{j+1}^z) \, \sigma_j^y \right] \right\} \, .
\end{equation*}
\end{widetext}

\ni The term responsible for the creation or annihilation of pairs of domain walls is suppressed under the assumption that the driving amplitude is chosen accordingly to $\mathcal{J}_0\left(\frac{A}{\Omega}\right) = 0$. The target model in Eq.\eqref{app:eq:hspin} is then obtained after the identification $\frac{\lambda}{2} \mathcal{J}_4\left(\frac{A}{\Omega}\right) = w$ and $\varphi = \frac{\pi \Phi}{2 L \Phi_0}$.

\section{Two-body problem}
\label{app:exactresults}

Consider the two body problem for the Hamiltonian \eqref{eq:app:model}. States in the physical Hilbert space are chosen as eigenstates of $\prod_j n_j$ and $\prod_j \sigma^z_{j+\frac{1}{2}}$. Due to the Gauss law, to every fermion configuration correspond two different spin orientations that differ for a flip of all the spins. Specializing to the two particles case, labeling as $j_1$ and $j_2$ the positions of the two fermions, with $j_1 < j_2$, the allowed states are the following:
\begin{align}
    \label{up}
    & \ket{j_1,j_2}_{\uparrow} = \ket{j_1,j_2}\ket{\downarrow_{1+ \frac{1}{2}} \dots \downarrow \, \uparrow_{j_1 + \frac{1}{2}} \dots \uparrow_{j_2 - \frac{1}{2}} \, \downarrow \dots \downarrow_{L+\frac{1}{2}}} \, , \\
    \label{down}
    & \ket{j_1,j_2}_{\downarrow} = \ket{j_1,j_2}\ket{\uparrow_{1+ \frac{1}{2}} \dots \uparrow \, \downarrow_{j_1 + \frac{1}{2}} \dots \downarrow_{j_2 - \frac{1}{2}} \, \uparrow \dots \uparrow_{L+\frac{1}{2}}} \, , 
\end{align}

\begin{figure*}[th]
\centering
\includegraphics[width = 0.7 \textwidth]{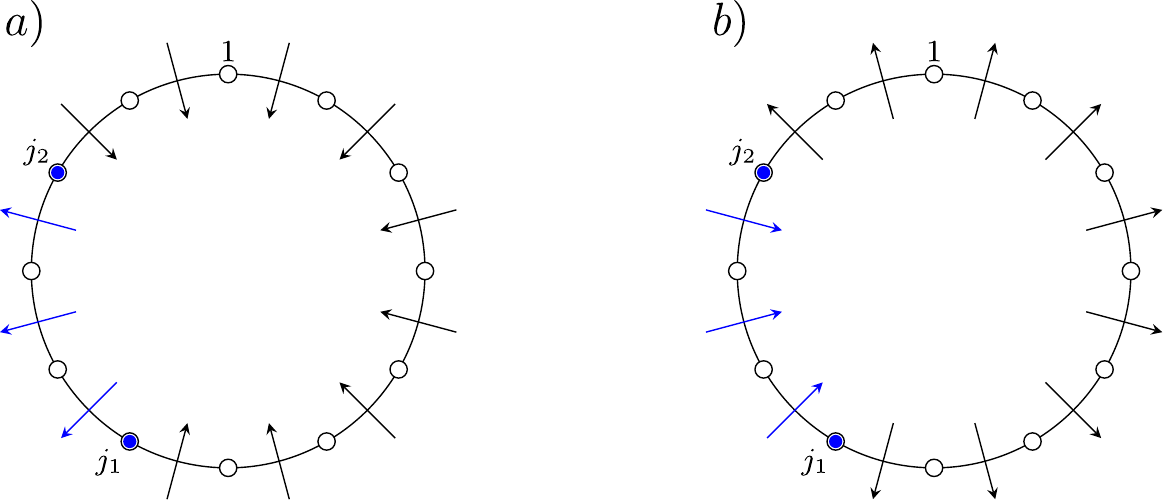}
\caption{\textit{Two particle states}. Here, the admissible two particle configurations are shown. The positive direction along the ring is taken to be clockwise. Outgoing and ingoing arrows represent up and down spins respectively. Blue arrows highlight the spins between the ordered fermion positions.(a) $\ket{j_1,j_2}_\uparrow$. (b) $\ket{j_1,j_2}_\downarrow$.}
\label{tpstates}
\end{figure*}

\ni where it is assumed that $s_{L+\frac{1}{2}} \equiv s_{1 - \frac{1}{2}}$ and
\begin{equation*}
    \ket{j_1,j_2} = c^\dagger_{j_1} c^\dagger_{j_2} \ket{0} \, . 
\end{equation*}

\ni The labels $\uparrow$ and $\downarrow$ refer to the orientation of the spins comprised between the ordered positions of the fermions and the corresponding states are unanbigously distinguished by the value of $\sigma^z_{L+\frac{1}{2}}$. See Figure \ref{tpstates} for a pictorial representation. 

\ni Every state in the two-particle space can thus be expanded in this basis as:
\begin{equation}
\label{expansion}
    \ket{\psi} = \sum_{j_1 < j_2} \sum_{a = \uparrow,\downarrow} \psi_a(j_1,j_2) \ket{j_1,j_2}_a \, .
\end{equation}

\ni For the sake of keeping translational invariance manifest in the calculations, it is convenient to adopt a relabeling of these states. In particular, choosing the clockwise direction along the ring as positive for definiteness, let $R$ be the length (in units of the lattice spacing) of a string of spins pointing in the "up" direction and originating from site $j$. Then, the expansion \eqref{expansion} can be rewritten as:
\begin{equation}
    \ket{\psi} = \sum_{j = 1}^L \sum_{R = 1}^{L-1} \psi(j,R) \ket{j,R} \, .
\end{equation}

\ni In particular, the two states in equations \eqref{up} and \eqref{down}, become:
\begin{align}
\label{jr1}
    & \ket{j_1,j_2}_\uparrow = \ket{j_1,r} \, ,\\
\label{jr2}
    & \ket{j_1,j_2}_\downarrow = \ket{j_2,L-r} \, ,
\end{align}

\ni where $r = j_2 - j_1$. In the basis $\{\ket{j,R}\}$, the two-body Hamiltonian can be written as follows:
\begin{widetext}
\begin{align}
\label{app:2bham}
\nonumber
    \mathcal{H}_{2b} = \sum_{j,R} & \Bigl\{ \tau \left( R- L/2\right) \proj{j,R} + w \Bigl[ e^{i \frac{2\pi \Phi}{L \Phi_0}} \left[(1-2 \, \delta_{j,1}) \ket{(j-1) \Mod{L},R+1}+ (1-2 \, \delta_{j,L-R+1}) \ket{j,R-1}\right] \\ 
    &+ e^{- i \frac{2\pi \Phi}{L \Phi_0}} \left[(1-2 \, \delta_{j,L}) \ket{(j+1) \Mod{L},R-1} + (1-2 \, \delta_{j,L-R}) \ket{j,R+1} \right] \Bigr]\bra{j,R} \Bigr\} \, .
\end{align}
\end{widetext}

\ni The $\delta$ factors in the previous expression account for the sign change as the particles hop between the $1^{st}$ and the $L^{th}$ sites. Notice that, if $R=1$ or $R=L-1$, two terms of the previous equation must be dropped, i.e. those relating $R=1$ with $R = 0$ and $R=L-1$ with $R=L$. This is equivalent to the Pauli exclusion principle, as $R=0,L$ correspond to fermions occupying the same site.

We wish to exploit the translational symmetry of $\mathcal{H}$. Denoted the translation operator as $\mathcal{T}$, its action on the states $\ket{j,R}$ results in $\mathcal{T} \ket{j,R} = (1 - 2\,\delta_{j+R,L} - 2\,\delta_{j,L})\ket{(j+1) \Mod{L},R}$, where the prefactor accounts for the minus sign we get upon acting with $\mathcal{T}$ on states with one of the two particles at the first or last site of the ring. In fact, in terms of the basis $\{\ket{j_1,j_2}_a\}$, this is equivalent to $\mathcal{T} \ket{j_1,L}_{\uparrow/\downarrow} = - \ket{1,j_1 + 1}_{\downarrow/\uparrow}$, and the minus sign arises as we rearrange the particles' positions in increasing order, due to anticommutation. The translation operator does not change the length of the string $R$, but it just shifts its origin along the ring. Thus, $\mathcal{T}$ can be diagonalized in each subspace of fixed $R$, where it has the matrix form
\begin{widetext}
\begin{equation}
    \mathcal{T}^{(R)}_{j',j} = \bra{j',R} \mathcal{T} \ket{j,R} =
    \begin{tikzpicture}[baseline=(current bounding box.center)]
        \matrix(T) [matrix of math nodes, nodes in empty cells, right delimiter = {)},left delimiter = {(}]{
         0 &   &   &   &   &   &   &   &   &   &   & \phantom{1}  & \llap{-}1  \\
         1 & 0 &   &   &   &   &   &   &   &   &   &   &  \\
           & 1 & 0 &   &   &   &   &   &   &   &   &   &  \\
           &   & 1 & 0 &   &   &   &   &   &   &   &   &  \\
            &  &   &   &   &   &   &   &   &   &   &   &   \\
           &  &   &   &   &   &   &   &   &   &   &   &   \\
           &  &   &   &   & 1 & 0 &   &   &   &   &   &   \\
         &  &   &   &   &   & \llap{-}1 & 0 &   &   &   &   &  \\
          &  &   &   &   &   &   & 1 & 0 &   &   &   &   \\
          &  &   &   &   &   &   &   &   &   &   &   &   \\
          &  &   &   &   &   &   &   &   &   &   &   &   \\ 
          &  &   &   &   &   &   &   &   &   & 1 & 0 &   \\
          &  &   &   &   &   &   &   &   &   &   & 1 & 0 \\
    };
        \draw[dotted] (T-1-1) -- (T-1-12);
        \draw[dotted] (T-1-13) -- (T-13-13);
        \draw[dotted] (T-4-4) -- (T-7-7);
        \draw[dotted](T-9-9) -- (T-12-12);
        \draw[dotted](T-4-3) -- (T-7-6);
        \draw[dotted](T-9-8) -- (T-12-11);
        \draw[-stealth] (T-8-7) -- (T-13-7);
        \path (-0.13,-2.7) node[] {$(L-R)^{th}$ column} coordinate(P);
    \end{tikzpicture} \, .
\end{equation}
\end{widetext}

\ni It is straightforward to check that the eigenvalues of $\mathcal{T}^{(R)}$ satisfy
\begin{equation}
    \lambda^L = 1 \implies \lambda = e^{-iK}, \quad K = \frac{2\pi n}{L} \, , \,  n = 0,\dots,L-1 
\end{equation}

\ni and its normalized eigenvectors, that we label as $\ket{K,R}$, are given by:
\begin{equation}
\label{kjr}
    \ket{K,R} = \frac{1}{\sqrt{L}} \sum_{j=1}^L e^{iKj} \, \theta_{L-R}(j) \ket{j,R} \, ,
\end{equation}

\ni where we defined
\begin{equation}
    \theta_{L-R}(j) = \begin{cases}
                            \ 1 \quad,\quad j\leq L-R \\
                            \ \llap{-} 1 \quad,\quad j>L-R
    \end{cases} \, .
\end{equation}

Any state of the two particle subspace can now be expanded in terms of this basis as
\begin{equation}
\label{kexp}
    \ket{\psi} = \sum_K \sum_R \psi(K,R) \ket{K,R} \, .
\end{equation}

\ni Now, since $\commutator{\mathcal{H}_{2b}}{\mathcal{T}} = 0$, then $\mathcal{H}_{2b}$ is block-diagonal in a basis of $\mathcal{T}$. Each block of $\mathcal{H}_{2b}$ will correspond to a fixed value of $K$ and thus will connect all the states $\ket{K,R}$, with $R$ spanning the set $\{1,\dots,L-1\}$. We denote these blocks as $\mathcal{H}^K_{2b}$.
Hence, we are interested in evaluating $\mathcal{H}_{2b} \ket{K,R}$, which can be deduced from Eq.\eqref{app:2bham}:
\begin{widetext}
\begin{align*}
    \mathcal{H}_{2b} &\ket{j,R} = w \Bigl[ e^{i \frac{2\pi \Phi}{L \Phi_0}} \left[(1-2 \, \delta_{j,1}) \ket{(j-1) \Mod{L},R+1}+ (1-2 \, \delta_{j,L-R+1}) \ket{j,R-1}\right] \\ 
    &+ e^{- i \frac{2\pi \Phi}{L \Phi_0}} \left[(1-2 \, \delta_{j,L}) \ket{(j+1) \Mod{L},R-1} + (1-2 \, \delta_{j,L-R}) \ket{j,R+1} \right] \Bigr] + \tau \left( R-\frac{L}{2} \right) \ket{j,R}\, .
\end{align*}
\end{widetext}

Now, we observe that:
\begin{widetext}
\begin{align*}
    &\sum_{j=1}^L e^{iKj} \theta_{L-R}(j) (1-2 \, \delta_{j,1}) \ket{(j-1)\Mod{L},R+1} = \sum_{j=2}^L e^{iKj} \theta_{L-R}(j) \ket{j-1,R+1} - e^{iK} \ket{L,R+1} \\ 
    &= \sum_{j=1}^{L-1} e^{iK(j+1)} \theta_{L-R}(j+1) \ket{j,R+1} - e^{iK}\ket{L,R+1} = e^{iK} \sum_{j=1}^L e^{iKj} \theta_{L-R-1}(j) \ket{j,R+1} = \sqrt{L} \, e^{iK} \ket{K,R+1}\, ,
\end{align*}
\end{widetext}

\ni where, in the last equality, we used the property that $\theta_{L-R}(j \pm 1) = \theta_{L-R \mp 1}(j)$. In the same way, we can prove that $\frac{1}{\sqrt{L}} \sum_{j=1}^L e^{iKj} \theta_{L-R}(j) (1-2 \, \delta_{j,L}) \ket{(j+1) \Mod{L},R-1} = e^{-iK} \ket{K,R-1}$. Moreover, noticing that
\begin{align*}
     (1-2 \, \delta_{j,L-R+1}) \, \theta_{L-R}(j) &= \theta_{L-R+1}(j) \, ,\\
     (1-2 \, \delta_{j,L-R}) \, \theta_{L-R}(j) &= \theta_{L-R-1}(j) \, , 
\end{align*}

\ni we easily obtain
\begin{align*}
    &\frac{1}{\sqrt{L}} \sum_{j=1}^L e^{iKj} \theta_{L-R}(j) (1-2 \, \delta_{j,L-R+1}) \ket{j,R-1} = \ket{K,R-1} \, , \\
    &\frac{1}{\sqrt{L}} \sum_{j=1}^L e^{iKj} \theta_{L-R}(j) (1-2 \, \delta_{j,L-R}) \ket{j,R+1} = \ket{K,R+1} \, . 
\end{align*}

\ni We can finally evaluate the action of the Hamiltonian on the eigenstates of $\mathcal{T}$. We promptly find:
\begin{align}
\nonumber
    &\mathcal{H}_{2b} \ket{K,R} = 2 \, w \cos\left({\frac{K}{2}} + \frac{2 \pi \Phi}{L \Phi_0} \right) \Bigl\{ e^{i\frac{K}{2}} \ket{K,R+1}\\
    &+ e^{-i \frac{K}{2}} \ket{K,R-1} \Bigr\} + \tau \left(R - \frac{L}{2} \right) \ket{K,r} \,.
\end{align}

Now, let $\ket{\Psi_K}$ be an eigenstate of a block $\mathcal{H}^K_{2b}$ of $\mathcal{H}_{2b}$ associated with an energy $E$:
\begin{equation}
\label{eig}
    \mathcal{H}^K_{2b} \ket{\Psi_K} = E \ket{\Psi_K} \, .
\end{equation}

\ni Since $K$ is held fixed, $\ket{\Psi_K}$ can be expanded as:
\begin{equation}
    \ket{\Psi_K} = \sum_R \Psi_K(R) \ket{K,R} \, .
\end{equation}
We evaluate $\bra{K,R} \mathcal{H}_{2b}^K \ket{\Psi_K} = E \, \Psi_K(R)$. The eigenvalue equation \eqref{eig} translates in a difference equation for the wave function. Defining $w_K(\Phi) = 2 \, w \cos \left( \frac{k}{2} + \frac{2 \pi \Phi}{L \Phi_0} \right)$ and shifting $E$ by the energy of the negatively polarized vacuum state $-\frac{\tau L}{2}$, we get:
\begin{align}
\label{diff}
\nonumber
    w_K(\Phi) \Bigl\{ e^{-i \frac{K}{2}} &\Psi_K(R+1) + e^{i \frac{K}{2}} \Psi_K(R-1) \Bigr\} \\ & \quad\quad + \tau R \, \Psi_K(R) = E \, \Psi_K(R) \, .
\end{align}

\ni Choosing the ansatz
\begin{equation}
    \Psi_K(R) = e^{iK \frac{R}{2}} \, \phi_K(R) \, , 
\end{equation}

\ni equation \eqref{diff} becomes
\begin{equation}
    \phi_K(R+1) + \phi_K(R-1) = \frac{\tau}{w_K} \left(\frac{E}{\tau} - R\right) \phi_K(R) \, . 
\end{equation}

\ni This is just the recursion relation of Bessel functions, i.e. $f_{\nu + 1}(x) + f_{\nu -1}(x) = \frac{2 \, \nu}{x} f_n(x)$, which leads to the solution:
\begin{equation}
    \phi_K(R) = \alpha \, \mathcal{J}_{\frac{E}{\tau} - R} \left(\frac{2 \, w_K(\Phi)}{\tau} \right) + \beta \, \mathcal{Y}_{\frac{E}{\tau} - R} \left(\frac{2 \, w_K(\Phi)}{\tau} \right) \, ,
\end{equation}

\ni $\mathcal{J}$ and $\mathcal{Y}$ respectively being Bessel functions of the first and second kind. We still have to impose the boundary conditions
\begin{equation}
    \phi_K(R) = 0 \quad , \quad R=0,L \, ,
\end{equation}

\ni that we have already noticed are equivalent to Pauli exclusion principle. $\phi_K(0) = 0 $ results in
\begin{equation}
\label{sol}
    \phi_K(R) = \mathcal{N} \left\{ \frac{\mathcal{J}_{\frac{E}{\tau} - R}\left(\frac{2 \, w_K(\Phi)}{\tau}\right)}{\mathcal{J}_{\frac{E}{\tau}}\left(\frac{2 \, w_K(\Phi)}{\tau}\right)} - \frac{\mathcal{Y}_{\frac{E}{\tau} - R}\left(\frac{2 \, w_K(\Phi)}{\tau}\right)}{\mathcal{Y}_{\frac{E}{\tau}}\left(\frac{2 \, w_K(\Phi)}{\tau}\right)} \right\} \, ,
\end{equation}

\ni while the remaining condition $\phi_K(L) = 0$ fixes $E$ to those values that solve
\begin{equation}
\label{eco}
    \frac{\mathcal{J}_{\frac{E}{\tau} -L}\left(\frac{2 \, w_K(\Phi)}{\tau}\right)}{\mathcal{J}_{\frac{E}{\tau}}\left(\frac{2 \, w_K(\Phi)}{\tau}\right)} - \frac{\mathcal{Y}_{\frac{E}{\tau} - L}\left(\frac{2 \, w_K(\Phi)}{\tau}\right)}{\mathcal{Y}_{\frac{E}{\tau}}\left(\frac{2 \, w_K(\Phi)}{\tau}\right)} = 0 \, .
\end{equation}

\ni The eigenenergies will thus be fixed by an index $l$, which denotes the $l-th$ solution of Eq.\eqref{eco} and by the center of mass momentum $K$. The state $\ket{\Psi_{K,l}}$ associated to one energy eigenvalue, will then have the expansion:
\begin{equation}
    \ket{\Psi_{K,l}} = \sum_R e^{iK \frac{R}{2}} \, \Phi_{K,l}(R) \ket{K,R} \, .
\end{equation}

We may now rewrite these results in terms of the original basis $\left\{\ket{j_1,j_2}_{\uparrow,\downarrow}\right\}$. In order to write the expressions for the wave functions $\Psi_{K,l}^\uparrow(j_1,j_2)$ and $\Psi_{K,l}^\downarrow(j_1,j_2)$, we need to evaluate ${}_{\uparrow,\downarrow} \! \braket{j_1,j_2}{K,R}$. Equations \eqref{jr1} and \eqref{jr2} yield:
\begin{align}
    & {}_{\uparrow} \! \braket{j_1,j_2}{j,R} = \delta_{j,j_1} \, \delta_{R,j_2 - j_1} \, , \\
    & {}_{\downarrow} \! \braket{j_1,j_2}{j,R} = \delta_{j,j_2} \, \delta_{R,L-j_2+j_1} \, .
\end{align}
Finally, from the expansion of the states $\ket{K,R}$ in terms of $\ket{j,R}$, Eq.\eqref{kjr}, we find:
\begin{align*}
    {}_{\uparrow} \! \braket{j_1,j_2}{K,R} &= \frac{1}{\sqrt{L}} \, e^{iK j_1} \, \theta_{L-R}(j_1) \,\delta_{R,j_2 - j_1} \\ &= \frac{1}{\sqrt{L}} \, e^{iK j_1} \, \theta_{L - j_2}(0) \, \delta_{R,j_2 - j_1} \, ,
\end{align*}
\begin{align*}
    {}_{\downarrow} \! \braket{j_1,j_2}{K,R} &= \frac{1}{\sqrt{L}} \, e^{iK j_2} \, \theta_{L-R}(j_2) \,\delta_{R,L -j_2 +j_1} \\ &= \frac{1}{\sqrt{L}} \, e^{iK j_2} \, \theta_{-j_1}(0) \, \delta_{R,L-j_2+j_1} \, ,
\end{align*}

\ni where we used the fact that $\theta_{L-j_2 + j_1}(j_1) = \theta_{L-j_2}(0)$ and $\theta_{j_2 - j_1}(j_2) = \theta_{-j_1}(0)$. Moreover, since $j_1,j_2 \in \{1,\dots,L\}$, $\theta_{L-j_2}(0) = 1$ and $\theta_{-j_1}(0) = -1$. Hence:
\begin{align}
    & {}_{\uparrow} \! \braket{j_1,j_2}{K,R} = \frac{1}{\sqrt{L}} \, e^{iK j_1} \, \delta_{R,j_2 - j_1} \, ,\\
    & {}_{\downarrow} \! \braket{j_1,j_2}{K,R} = - \frac{1}{\sqrt{L}} \, e^{iK j_2} \, \delta_{R,L-j_2+j_1} \, .
\end{align}

\ni Therefore, we can now compute the wave functions:
\begin{align}
    & \Psi_{K,l}^\uparrow(j_1,j_2) = \frac{1}{\sqrt{L}} \, e^{i K \frac{j_1 + j_2}{2}} \, \phi_{K,l}(j_2-j_1) \\ \, 
    & \Psi_{K,l}^\downarrow(j_1,j_2) = - \frac{1}{\sqrt{L}} \, e^{iK \frac{j_1 + j_2 + L}{2}} \, \phi_{K,l}(L - j_2 + j_1) \, .
\end{align}

\ni In terms of the center of mass and the relative coordinates, namely $s = \frac{j_1 + j_2}{2}$ and $r = j_2 - j_1 > 0$, we have:
\begin{align}
    & \Psi_{K,l}^\uparrow(s,r) = \frac{1}{\sqrt{L}} \, e^{i K s} \, \phi_{K,l} (r) \, , \\
    & \Psi_{K,l}^\downarrow(s,r) = - \frac{1}{\sqrt{L}} \, e^{iK \left(s +\frac{L}{2}\right)} \, \phi_{K,l} (L - r) \, . 
\end{align}

\section{Two-body current and fluctuations in an energy eigenstate}
\label{app:current_and_fluctuations}
\begin{figure}[th!]
    \centering
    \includegraphics[width = \linewidth]{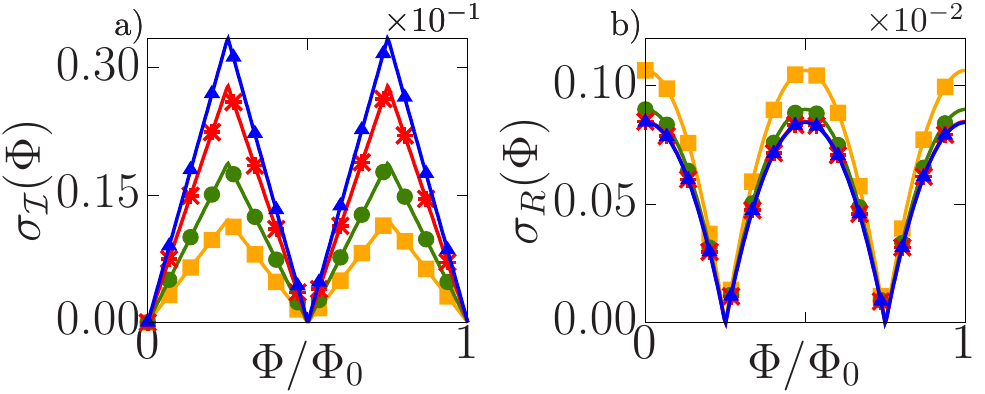}
    \caption{\textit{Current and string length fluctuations}. In panel (a) we plot the fluctuations of the single meson current; in panel (b), the fluctuations of the meson extent. For visualization purposes, all the curves in panel (b) have been shifted to zero minimum. The fluctuations are plot for $\tau = 0.5,1,2,3$, respectively in yellow, green, red and blue. All the data are shown for $L=21$.}
    \label{fig:fluc}
\end{figure}

For $\tau = 0$, $\commutator{\mathcal{H}}{\mathcal{I}} = 0$. Therefore, in any energy eigenstate  $\sigma_\mathcal{I}^2 = \mean{\mathcal{I}}^2 - \mean{\mathcal{I}}^2 = 0$. This is not true anymore for $\tau \neq 0$, leading to $\sigma_\mathcal{I} \neq 0$.
Direct analysis of the equations of motion for the two-particle problem provide that, in any energy eigenstate of fixed center of mass momentum $K$, the current and its fluctuations are expressed as
\begin{align}
    \mean{\mathcal{I}}_{K,l} &= g(K,\Phi) (E_{K,l} - \tau \langle\hat{R}\rangle_{K,l}) \, , \\
    (\sigma_\mathcal{I})_{K,l} &= \tau \abs{g(K,\Phi)} (\sigma_R)_{K,l} \, ,
\end{align}
where $g(K,\Phi) = \frac{2 \pi}{L \Phi_0} \tan{\Bigl(\frac{K}{2} + \frac{2 \pi \Phi}{L \Phi_0}\Bigr)}$ and $\hat{R}$ is the operator associated to the length of a string of positive spins. Therefore, we conclude that  the current fluctuation calculated in each energy eigenstate is related to the fluctuation of the spatial scale on which the meson can be confined - see Fig.\ref{fig:fluc}.

\section{Ground state current}
\label{app:ground_current}
\begin{figure}[th!]
    \centering
    \includegraphics[width = \linewidth]{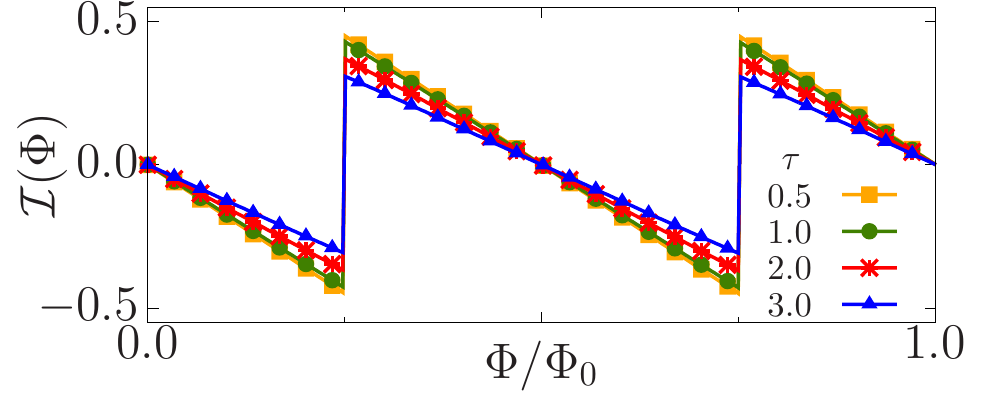}
    \caption{\textit{Ground state current}. We show $\langle \mathcal{I}(\Phi) \rangle $ in the confined regime. Here, we considered $L = 12$ and $N = 8$ particles.}
    \label{fig:sawtooth}
\end{figure}
The persistent current is given by the ground state expectation value of the operator
\begin{equation}
    I(\Phi) = -i \frac{2 \pi w}{L \Phi_0}  \sum_j (e^{i \frac{2\pi}{L} \Phi/\Phi_0} c^\dagger_j \sigma_{j+\frac{1}{2}}^x c_{j+1} - h.c.) \, .
\end{equation}

\ni For any value of the string tension $\tau$, the current displays a saw-tooth behaviour with period of half flux quantum $\Phi_0/2$ - see Fig.\ref{fig:sawtooth}. In the deconfined regime $\tau = 0$, the Hamiltonian (1) of the main text is block diagonal in a basis of eigenvectors of the Wilson loop operator $W = \prod_j \sigma_{j\frac{1}{2}}^x$. This can be made manifest by writing 
\begin{equation}
\label{app:free}
    \mathcal{H} = w \left[ \sum_{j = 1}^{L-1} (d^\dagger_j d_{j+1} + h.c.)
    + W (e^{i \, \frac{2\pi \Phi}{\Phi_0}} d_L^\dagger d_1 + h.c.) \right] \, ,
\end{equation}
where  $d_j = \prod_{k = 1}^{j-1} \sigma_{k + \frac{1}{2}}^x \, c_j$. 
Each block of the Hamiltonian hence describes free fermions with twisted boundary conditions, and has eigenvalues $E = 2w \sum_{i=1}^N cos (k_i + \varphi/L)$, with $N$ and $k_i$ being the number of particles and their quasi-momenta, respectively. Due to the original gauge constraints on periodic boundary conditions, the parity of $N$ is fixed to be even. The total Aharonov-Bohm phase $\varphi$ acquired by the particles when going around the loop is
$ 
\varphi = \frac{2\pi \Phi}{\Phi_0} +arg (W)  \, ,
$ 
where $W$ is the eigenvalue of the Wilson loop operator. Therefore $arg(W)=2n \pi$ or $arg(W)=(2n+1) \pi$, $n$ being integer,  for $W=1$ or $W=-1$ respectively.
Finally, by inspection of the spectra of the two blocks,  $\Phi/\Phi_0 = 1/4,3/4$ correspond to an energy degeneracy, in turn providing the  reduction of  the current flux periodicity.

In the confined regime $\tau \neq 0$, we claimed that the reduced period of the current is determined by the formation of confined bound states. Relevant insights into this case can be obtained by a perturbative analysis that can be carried out for $\tau \gg w$. Indeed, a second order Schrieffer-Wolf transformation leads to the effective model \cite{borla2020confined}
\begin{align}
\label{eq:perturbative}
    \mathcal{H}_{\textit{eff}} = P^\dagger \sum_j \bigl[ - \bar{w} (e^{i \frac{4 \pi}{L} \Phi/\Phi_0} b_j^\dagger b_{j+1} + h.c. ) + \, \bar{V} \, n^b_j n^b_{j+2} \bigr] P \, ,
\end{align}
in which $b_j^\dagger = c^\dagger_j \sigma_{j+\frac{1}{2}}^x c_{j+1}^\dagger$, are hard-core bosonic operators hopping with a rate  $\bar{w} = \frac{w^2}{\tau}$,  and  interacting through $\bar{V} = 2 \bar{w}$. The operator $P$ projects $H_{\textit{eff}}$ onto the Hilbert space of those states that satisfy the constraint $b^\dagger_j b^\dagger_{j+1} = 0$. We note that bosons dynamics occurs as second order processes,  in which pairs of fermions move together coherently. Strikingly, the effective magnetic field is  effectively doubled in such a way one flux quantum is shared by two particles. This feature directly implies  that the periodicity of the current is effectively halved.

\begin{figure}
    \centering
    \includegraphics[width = \linewidth]{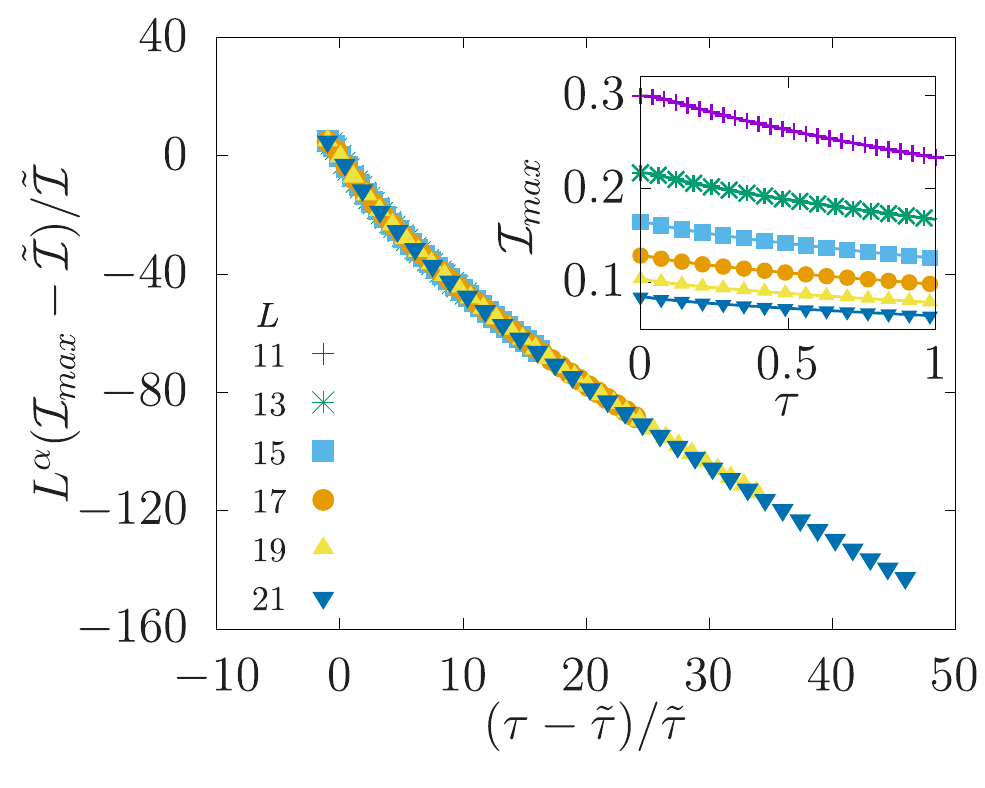}
    \caption{\textit{Scaling of the ground state current in the two particle sector}. Here we display the amplitude of the ground sate single meson current. The scaling points $(\tilde{\tau}(L),\tilde{\mathcal{I}}(L))$ correspond to the inflections of $\mathcal{I}_{max}(L)$. $\alpha = 2.08$, $\tilde{\tau}(L) \sim L^{-2.98}$, $\tilde{\mathcal{I}}(L) \sim L^{-1.91}$.}
    \label{app:fig:scaling}
\end{figure}

\begin{figure*}[th!]
    \centering
    \includegraphics[width = 0.85 \linewidth]{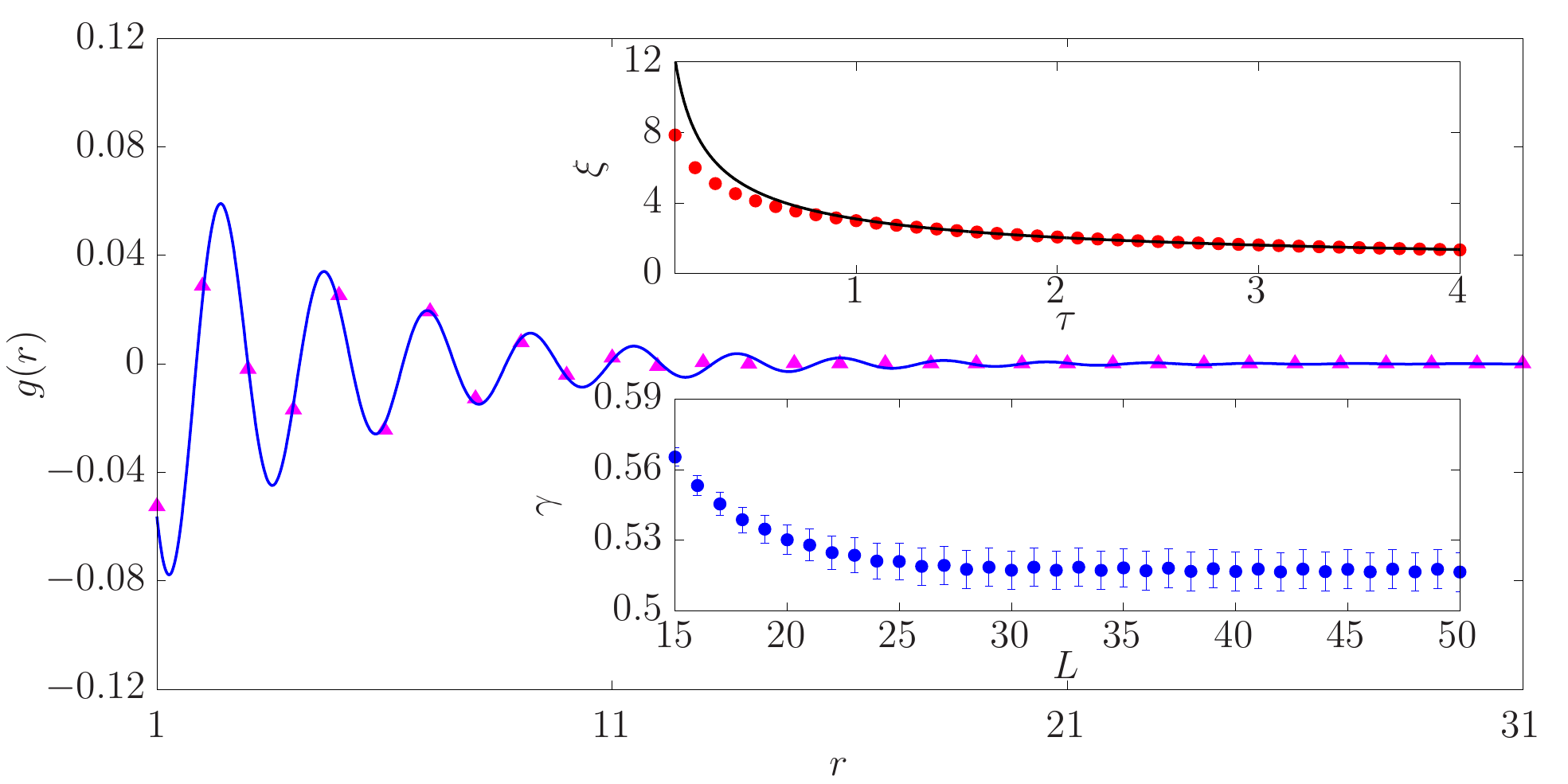}
    \caption{\textit{Ground state single meson correlator}. The main figure shows the raw data (points) and the fitting function (curve) for $g(r)$ at $L=32$ and $\tau = 0.5$. Upper inset: the points correspond to the values of the correlation length $\xi$ as a function of $\tau$, for $L = 32$, while the curve shows the fitting function for $\xi(\tau)$; the error bars are not displayed because they would not be visible. Lower inset: decay of the exponent $\gamma$ with increasing system size $L$.} 
    \label{fig:correlator}
\end{figure*}

In the two-particles sector, we studied the maximum of the ground state current $\mathcal{I}_{\text{max}}$ as a function of the string tension $\tau$ and of the number of sites in the ring $L$. We find that $\mathcal{I}_\text{max}$ displays scaling behaviour, as shown in Fig.\ref{app:fig:scaling}. For
large enough $\tau$ , the scaling function provides the behaviour $\mathcal{I}_\text{max} \sim L^{-2}$, from which it deviates substantially in the regime of small $\tau$.

\section{Gauge-invariant correlator}
\label{app:correlator_section}

We present the analysis of the exponential decay of the gauge invariant correlator $g(r) = \langle c^\dagger_{j+r} \prod_{l = j}^{j+r-1} \sigma^x_{l+\frac{1}{2}} c_j \rangle$. We fit the data obtained via exact diagonalization, with the law $g(r) = p_1 e^{-r/p_2} \cos(p_3 r + p_4)$.
Hence, we extract the correlation length $\xi$ from the fit parameter $p_2$. We find that the correlation length decays with the string tension $\tau$, with the law: $\xi(\tau) = a \, \tau^{-\gamma}$. The exponent $\gamma$ is a weakly decreasing function of the system size $L$ and is seen to saturate at a value $\gamma \sim 0.52$ for large $L$.

\section{Two-body dynamics}
\label{app:twobodydynamics}
The first-quantized two-body Hamiltonian, Eq.\eqref{app:2bham}, can be rewritten in terms of suitably defined canonically conjugate operators. Denoting $s$ and $R$ as the center of mass coordinate and the string length of a two particle configuration, as described in the previous section, we introduce $\hat{Q} = \sum_s s \proj{s}$, $\hat{R} = \sum_R R \proj{R}$, $e^{-i \hat{K}/2} = \sum_s \ketbra{s+1/2}{s}$, $e^{-i\hat{p}} = \sum_R \ketbra{R+1}{R}$. These operators define the center of mass position, the string length, the center of mass momentum and the relative momentum, respectively. They satisfy canonical commutations relations $\commutator{\hat{Q}}{\hat{K}} = \commutator{\hat{R}}{\hat{p}} = i$. It is implicit that summations over $s$ and $R$ are indeed constrained by the lattice: to odd/even values of the string length $R$, can only correspond half-integer/integer values of the center of mass coordinate $s$. The two-body Hamiltonian is thus rewritten as:
\begin{equation}
    \mathcal{\hat{H}}_{2b} = 2 \, w \, cos\left( \frac{\hat{K}}{2} + \frac{2 \pi \Phi}{L \Phi_0} \right) cos(\hat{p}) + \frac{\tau}{2} \hat{R} \, .
\end{equation}

\ni The Pauli constraints at $R = 0,L$ are implicit. The Heisenberg equations of motion can be integrated exactly, away from the boundary conditions at $R=0,L$ \cite{collura2022time}. The string length is seen to perform Bloch oscillations, as
\begin{align}
\nonumber
        \hat{R}(t) &= \hat{R} - \frac{4w}{\tau} \cos\left( \frac{\hat{K}}{2} + \frac{2 \pi \Phi}{L \Phi_0} \right) \times \\ & \times \left[ \sin(\tau t) \, \sin(\hat{p}) - (1- \cos(\tau t)) \, \cos(\hat{p}) \right] 
\end{align}

\ni and $\hat{p}(t) = \hat{p} - \tau t$, with $\hat{p} = \hat{p}(t = 0)$. Due to the coupling between center of mass and relative coordinates, we find that the Bloch oscillations of the string length propagate to oscillations of the center of mass position, with a velocity which is fixed by the current operator
\begin{equation}
    \partial_t \hat{Q}(t) = \partial_t \hat{Q}(t = 0) - \frac{L \Phi_0}{4 \pi} \left[\hat{\mathcal{I}}(t) - \hat{\mathcal{I}}(t = 0)\right] \, .  
\end{equation}

\ni Indeed, the dynamics of the current is related to the dynamics of the string length, as
\begin{equation}
    \mathcal{\hat{I}}(t) = \mathcal{\hat{I}} -\frac{2 \pi \tau}{L \Phi_0} \tan{\left(\frac{\hat{K}}{2} + \frac{2\pi\Phi}{L\Phi_0}\right)} \left[\hat{R}(t) - \hat{R}\right] \, ,
\end{equation}

\ni where $\mathcal{\hat{I}} = \mathcal{\hat{I}}(t = 0)$.

In this paper, we considered the dynamics of a single meson after a quench of the external flux $\Phi: 0 \to \bar{\Phi}$. The initial state is chosen as the energy eigenstate, in the case $\Phi = 0$, corresponding to $K=0$, $l=1$, whose center of mass is localized within a Gaussian distribution - see Eq.(4) of the main text. Depending on the width of the Gaussian localization $\Sigma$, center of mass momenta are excited  that are different from $K = 0$. Due to the coupling between the center of mass and the relative motion, levels corresponding to $l \neq 1$ are also excited. Therefore, the initial state can be generally expanded as $\ket{\psi_0} = \sum_{K,l} c_{K,l} \ket{\Psi_{K,l}}$. In the case of a delta-like localization, $\Sigma \to 0$, a perturbative calculation to second order in $w/\tau$ provides
\begin{align}
\label{perturbative current}
\nonumber
    &\mean{\mathcal{I}(t)} = \frac{8 \pi w^2}{\tau L^2} \sum_K \sin\left[K + (4 \pi/L) \bar{\Phi}/\Phi_0\right] \times \\ & \times \cos\left[ \left(\tau + (4 w^2 /\tau) \cos^2[K/2 + (2 \pi/L) \bar{\Phi}/\Phi_0] \right) t\right] \, .
\end{align}

 \ni The last expression reproduces the features observed in Fig.5 of the main text.

\section{Quench dynamics}
\label{app:quench_section}

\begin{figure}[th!]
    \centering
    \includegraphics[width = \linewidth]{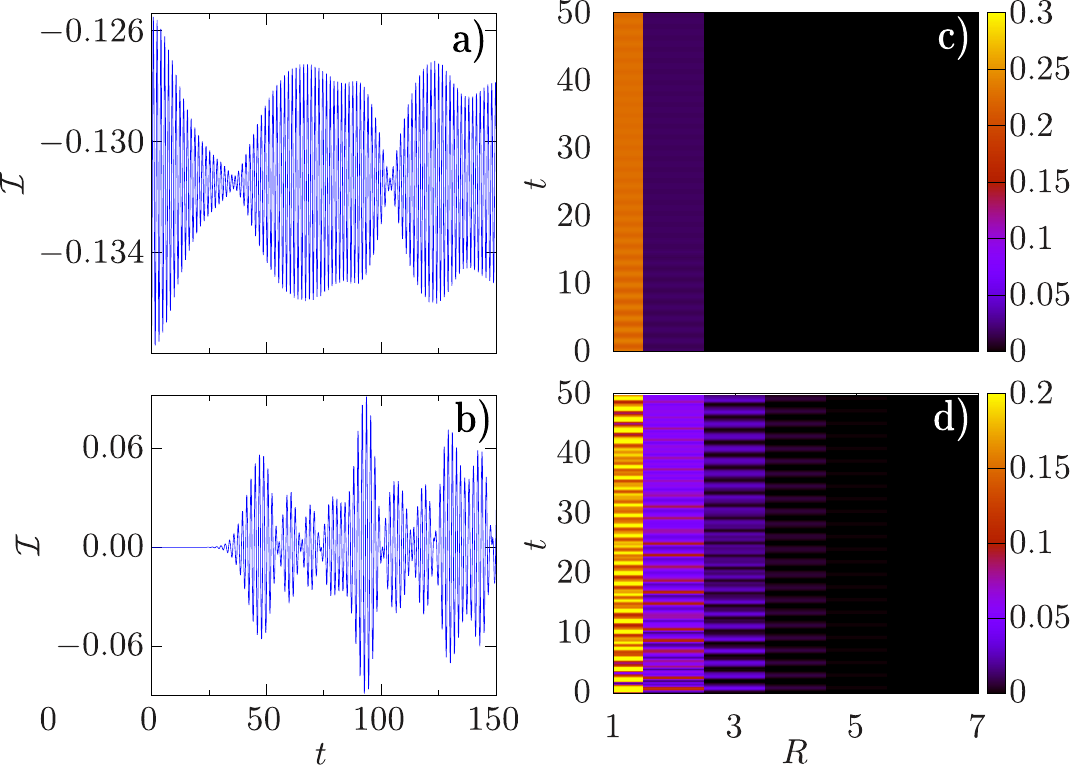}
    \caption{A single meson whose center of mass has been localized within a gaussian distribution of width $\Sigma$, is put in motion by quenching the flux to the value $\Phi/\Phi_0 = 0.8$. The average current (representing the center of mass velocity) and the probability to find the meson with string length $R$ (labelling the length of a string of spin ups connecting the two bound particles), $\mathcal{I}(t)$ and $P(R) = \sum_s |\braket{s,R}{\psi(t)}|^2$, are displayed in panels a),b) and c),d) respectively. The upper panels refer to $\Sigma = 2$, $\tau = 3$; the lower panels  refer to $\Sigma = 10^{-6}$, $\tau = 3$. In all cases, we consider $L = 21$ sites.}
\label{app:fig:quench}
\end{figure}

Here, we present further results to support our analysis of the quench dynamics. In Fig. \ref{app:fig:quench}, we plot the current and the string dynamics after a quench of the external flux, as a complement to Fig.(2) in the main text. In this case, a larger value of the string tension ($\tau = 3$) is considered, while the width of the gaussian localization $\Sigma$ is chosen as either $\Sigma = 2$ or $\Sigma \to 0$. As detailed in the main text, this choice is reflected in the behaviour of the current, displaying clear beating modes and a strongly delayed motion in the case $\Sigma \to 0$. On the other hand, the string oscillations are more suppressed and faster, as can be seen by comparison to the equivalent cases for $\tau = 1$, shown in Fig.(2) of the main text. 

\section{Single particle Aharonov-Bohm oscillations}
\label{app:ab_section}
\begin{figure}[th!]
    \centering
    \includegraphics[width = \linewidth]{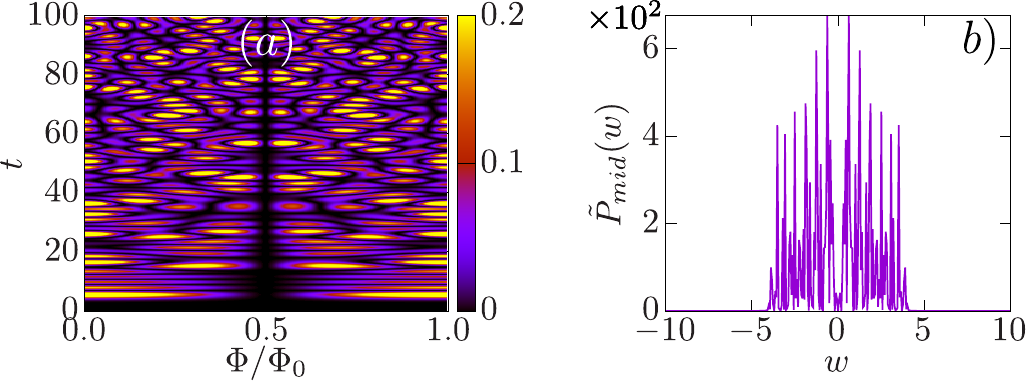}
    \caption{\textit{Single particle Aharonov-Bohm oscillations}. A single particle wave function, initialized in the site $j=j_0$ as specified in Eq.(\ref{app:initialstate}), evolves along the ring and is monitored at $j=j_0 + L/2$. Panel (a) displays $P_{mid}(t) = |\braket{j_0 + L/2}{\psi(t)}|^2 $. Panel (b) displays  the Fourier transform of $\left(P_{mid}(t) - \overline{P_{mid}}\right) = \tilde{P}_{mid}(w)$ at $\Phi = 0$. The figure refers to $L = 20$ sites and the hopping parameter is set to $h = 1$.}
    \label{fig:abfree}
\end{figure}
In this section, we present the analysis of the Aharonov-Bohm oscillations for a single particle in a ring pierced by a magnetic flux - see Fig.\ref{fig:abfree}, for comparison with the case of a confined meson. The particle is initially localized at site $j_0$
\begin{equation}
\label{app:initialstate}
    \ket{\psi(t=0)} = c^\dagger_{j_0} \ket{0}
\end{equation}

\ni and is evolved by means of the free-hamiltonian $\mathcal{H}_{free} = \sum_j h (e^{i \frac{2\pi\Phi}{L\Phi_0}} c^\dagger_j c_{j+1} + h.c.)$.
\ni We point out that, in this case, the analysis is carried out for $\Phi/\Phi_0 \in [0,1]$. Indeed, a free fermion system displays a periodic behaviour with a period of one flux quantum $\Phi_0$, whilst the model in Eq.(1) of the main text is characterized by an halved periodicity, fixed by $\Phi_0/2$. Direct calculation of $\mathcal{A}_{mid}(t) = \braket{j_0 + L/2}{\psi(t)}$ yields
\begin{align}
\nonumber
    \mathcal{A}_{mid}&(t) = 2 \sum_{\nu = 1}^\infty (-i)^{(2\nu -1)L/2} \times \\ & \times \mathcal{J}_{(2\nu -1)\frac{L}{2}}(2ht) \cos \left[(2\nu -1)\pi \frac{\Phi}{\Phi_0}\right] \, ,
\end{align}

\ni where $\mathcal{J}$ denotes the Bessel function of the first kind. The last expression provides that destructive interference occurs at all times if $\Phi = (2m+1) \frac{\Phi_0}{2}$, $m \in \mathbb{Z}$. 

A single meson, in the limit of very large string tension $\tau$, behaves like a free particle, hopping on the lattice with a renormalized amplitude - see Eq.\eqref{eq:perturbative}. The same calculation then yields
\begin{widetext}
\begin{equation}
    \mathcal{A}_{mid}^{meson}(t)  = 2 \sum_{\nu = 1}^\infty (-i)^{(2\nu -1)L/2} \mathcal{J}_{(2\nu -1)\frac{L}{2}}\left(\frac{2 w^2}{\tau}t\right) \cos \left[(2\nu -1)\pi \frac{2 \, \Phi}{\Phi_0}\right] \, ,
\end{equation}
\end{widetext}

\ni where $w$ is the hopping parameter in Eq.(1) of the main text. We observe that, in this case, destructive interference occurs at all times for $\Phi = (2m+1) \frac{\Phi_0}{4}$, $m \in \mathbb{Z}$ and that time is effectively rescaled by a factor of $w/\tau$.

\end{appendix}

%

\end{document}